\documentstyle[psfig,al6new]{article}
\newcommand{\be}{\begin{eqnarray}}
\newcommand{\ee}{\end{eqnarray}}

\title{Critical Exponent of Species-Size Distribution in Evolution}
\author{Chris Adami$^{1}$, Ryoichi Seki$^{1,2}$ \and Robel
  Yirdaw$^2$\\ \mbox{}\\ 
$^1$California Institute of Technology, Pasadena, CA
  91125 \\  $^2$California State University, Northridge, CA 91330}
\begin{document}
\maketitle
\begin{abstract}
  We analyze the geometry of the species-- and genotype-size
  distribution in evolving and adapting populations of single-stranded
  self-replicating genomes: here programs in the Avida world.  We find
  that a scale-free distribution (power law) emerges in complex
  landscapes that achieve a separation of two fundamental time scales:
  the relaxation time (time for population to return to equilibrium
  after a perturbation) and the time between mutations that produce
  fitter genotypes. The latter can be dialed by changing the mutation
  rate.  In the scaling regime, we determine the
  critical exponent of the distribution of sizes and strengths of
  avalanches in a system without coevolution, described by first-order
  phase transitions in single finite niches.
\end{abstract}
\section{Introduction}
Power law distributions in Nature usually signal the absence of a
scale in the region where the scaling is observed, and sometimes point
to critical dynamics. In Self-Organized-Criticality
(SOC)\nocite{BTW87,BTW88} (Bak, Tang, and Wiesenfeld 1987, 1988), for
example, power law distributions reveal the dynamics of an
unstable critical point, brought about by slow driving and a feed-back
mechanism between order parameter and critical parameter.
The critical dynamics is usually described within the language of
second-order phase transitions in condensed matter
systems~\nocite{SJD}(Sornette, Johansen and Dornic 1996),
but it can
be shown that SOC-type behavior also occurs within a dual description
in terms of the Landau-Ginzburg equation as {\em first-order}
transitions\nocite{GS} (Gil and Sornette 1996).
Indeed, it was shown that a power law distribution of {\em
  epoch-lengths}, that is, the time a particular species dominates the
dynamics of an adapting population, is explained by a
self-organized critical scenario~\nocite{CA2}(Adami 1995) that carries
the hallmark of
first-order phase transitions. Here, we measure the distribution of
abundances of {\em species} and genotypes in an artificial chemistry,
(the Avida Artificial Life system\nocite{AB1,OBA}, Adami and Brown
1995, Ofria, Brown and Adami 1998) and show
that the distribution is scale-free under a broad class of
circumstances, confirming the results reported in\nocite{CA2} Adami
(1995). 
In the next section, we discuss the first-order dynamics in more
detail and examine ``avalanches of invention'' from the point of view
of a thermodynamics of information. In Section III, we 
measure the critical exponent of the power law of genotype abundances
in the limit
of infinitesimal driving, i.e., infinitesimal mutation rate, and
discuss the role of the fitness landscape in shaping the
distribution. In Section IV, we repeat the analysis for a higher
taxonomic level (that of species) and discuss its relation to the
geometric distributions found by Burlando\nocite{BUR90,BUR93} (1990, 1993). 
Conclusions about
the evolutionary process drawn from the data  obtained in this paper
are presented in Section V.
\section{Self-Organization in Evolution}
The idea that the evolutionary process occurs in spurts, jumps, and bursts
rather than gradual, slow and continuous changes has been around for
over 75 years\nocite{WIL} (Willis 1922), 
but has gained prominence as ``punctuated
equilibrium'' through the work of Gould and
Eldredge\nocite{GE77,GE93} (1977, 1993). The
general idea is that evolutionary innovations are not bestowed upon an
existing species as a whole, gradually, but rather by the emergence of
{\em one} better adapted mutant which, by its superiority, serves as the
seed of a new breed that sweeps through an ecological niche  and
supplants the species previously occupying it. The global dynamics
thus has a microscopic origin, as shown experimentally, e.g., in populations
of {\it E. Coli} by Elena, Cooper and Lenski (1996).  

Such avalanches can be
viewed in two apparently contradictory ways. On the one hand we
may consider the wave of extinction touching all species that are connected
by their ecological relations, a process akin to percolation and
therefore suitably described by the language of second-order critical
phenomena\nocite{BS} (Bak and Sneppen 1993). Such a scenario relies on the 
{\em coevolution} of species (to build their ecological relations) and
successfully describes power-law distributions
obtained from the fossil record\nocite{SB96,BP96} (Sol\'e and
Bascompte 1996, Bak and Paczuski 1996). There is, on the other hand, a
description in terms of {\em informational} avalanches that does not
require coevolution and leads to the same statistics, as we show
here. Rather than contradicting the aforementioned
picture\nocite{NFST} (Newman et al. 1997), we believe it to be complementary. 

In the following, we set up a scenario in which {\em information} is
viewed as the agent of self-organization in evolving and adapting
populations. Information is, in the strict sense of Shannon theory, a
measure of correlation between two ensembles: here a population of
genomes and the environment it is adapting to. As described
elsewhere\nocite{IAL} (Adami 1998), 
this correlation grows as the population stores
more and more information about the environment via random
measurements, implementing a very effective {\em natural Maxwell
  demon}. Any time a stochastic event increases the information stored
in the population, a wave of extinction removes the less adapted
genomes and establishes a new era. Yet, information cannot leave the
population as a whole, which therefore may be thought of as protected
by a {\em semi-permeable membrane} for information, the hallmark of
the Maxwell demon. Let us consider this dynamics in more detail. 

The simple living systems we consider here are populations of
self-replicating strings of instructions, coded in an alphabet of
dimension ${\cal D}$ with variable string length $\ell$. The total
number of possible strings is exponentially large. Here, we consider
the subset of all strings currently in existence in a finite
population of size $N$, harboring $N_g$ different types, where $N_g\ll{\cal
  D}^\ell$. Each {\em genotype} (particular sequence of instructions) 
is characterized by its replication rate $\epsilon_i$, which depends 
on the sequence only, while its survival rate is given by
$\epsilon_i/\langle \epsilon\rangle$, in a ``stirred-reactor''
environment 
that allows a mean-field picture. This average replication rate 
$\langle \epsilon\rangle$ characterizes the fitness of the population
as a whole, and is given by
\be
\langle \epsilon\rangle=\sum_i^{N_g}\frac{n_i}N \epsilon_i\;,
\ee
where $n_i$ is the {\em occupation number}, or frequency, of genotype
$i$ in the population. As $N_g$ is not fixed in time, the average
depends on time also, and is to be taken over all genotypes currently
living. The total abundance, or size, of a genotype is then
\be
s_i=\int_0^\infty n_i(t)\, dt=\int_{T_c}^{T_e}n_i(t)\, dt\;,
\label{size}
\ee
where $T_c$ is the time of creation of this particular genotype, and
$T_e$ the moment of extinction. Before we obtain this distribution in
Avida, let us delve further into the statistical description of the
extinction events. 

\begin{figure}[t]
\caption{``Energies'' (inferiorities) of strings in a first-order
  phase transition with latent heat $\Delta\epsilon$.} 
\vskip 0.25cm
\centerline{\psfig{figure=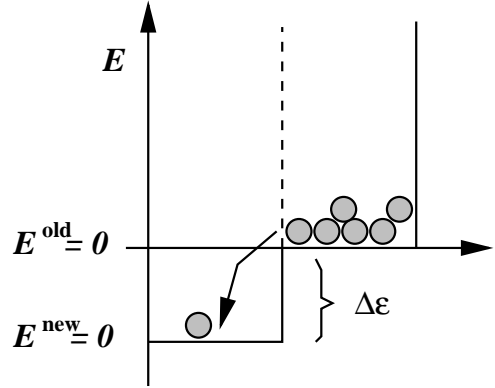,width=2.5in,angle=-90}}
\label{fig1}
\end{figure}

At any point in time, the fate of every string in the population is
determined by the craftiness of the best adapted member of the
population, described by $\epsilon_{\rm best}$. In this simple,
finite, world, which does not permit strings to affect other members
of the population except by replacing them, not being the best reduces
a string to an ephemeral existence. Thus, every string is
characterized by a {\em relative} fitness, or {\em inferiority}
\be
E_i=\epsilon_{\rm best}-\epsilon_i
\ee
which plays the role of an {\em energy} variable for strings of
information\nocite{IAL} (Adami 1998). Naturally, $\langle E\rangle=0$ characterizes
the {\em ground state}, or vacuum, of the population, and strings with
$E_i>0$ can be viewed as occupying {\em excited} states, soon to
``decay'' to the ground state (by being replaced by a string with
vanishing inferiority). Through such processes, the dynamics of the system
tend to minimize the average inferiority of the population, and the
fitness landscape of replication rates thus provides a Lyapunov
function. Consequently, we are allowed to proceed with our statistical
analysis. Imagine a population in equilibrium, at minimal 
average inferiority as allowed by the ``temperature'': the rate (or
more precisely, the probability) of
mutation. Imagine further that a mutation event produces a new
genotype, fitter than the others, exploiting the environment in novel
ways, replicating faster than all the others. It is thus endowed with a
new best replication rate, $\epsilon_{\rm best}^{\rm new}$, larger
than the old ``best'' by an amount $\Delta \epsilon$, and  redefining
what it means to be inferior. Indeed, all inferiorities must now be
{\em renormalized}: what passed as a ground state ($E=0$) string before now 
suddenly finds itself in an excited state. The seed of a new
generation has been sown, a phase transition must occur. In the
picture just described, this is a first-order phase transition with
latent heat $\Delta\epsilon$ (see Fig.~\ref{fig1}), starting at the 
``nucleation'' point, and leading to an expanding {\em bubble} of ``new
phase''. 
This bubble expands with a speed given by the Fisher velocity
\be
v\sim\sqrt{D\Delta\epsilon}\;, \label{eq4}
\ee
where $D$ is the diffusion coefficient (of information) in this
medium, until the entire population has been converted\nocite{CHU}
(Chu and Adami 1997).
This marks the end of the phase transition, as the population returns
to equilibrium via mutations acting on the new species, creating new
diversity and restoring the {\em entropy} of the population to its
previous value. This prepares the stage for a new avalanche, as only
an equilibrated population is vulnerable to even the smallest
perturbation. The system has returned to a critical point, driven by
mutations, self-organized by information. 

Thus we see how a first-order scenario, without coevolution, can lead
to self-organized and critical dynamics. It takes place within a
single, finite, ecological niche, and thus does not contradict the
dynamics taking place for populations that span many niches. Rather,
we must conclude that the descriptions complement each other, from the
single-niche level to the ecological web. Let us now take a closer look
at the statistics of avalanches in this model, i.e., at the
distribution of genotype sizes. 
\section{Exponents and Power Laws}

The size of an avalanche in this particular system can be approximated
by the size $s$ of the genotype that gave rise to it, Eq.~(\ref{size}).
We shall measure the distribution of these sizes $P(s)$ in the
Artificial Life system Avida, which implements a population of
self-replicating computer programs written in a simple machine
language-like instruction set of ${\cal D}=24$ instructions, with
programs of varying sequence length. In the course of
self-replication, these programs produce mutant off-spring because the
{\tt copy} instruction they use is flawed at a rate $R$ errors per
instruction copied, and adapt to an environment in which the
performance of {\em logical} computations on externally provided
numbers is akin to the catalysis of chemical
reactions\nocite{OBA} (Ofria, Brown and Adami 1998). In
this {\em artificial chemistry} therefore, successful computations
accelerate the metabolism (i.e., the CPU) of those strings that carry
the {\em gene} (code) necessary to perform the trick, and any program
discovering a new trick is the seed of another avalanche.

Avida is not a stirred-reactor environment (although one can be
simulated). Rather, the programs live on a two-dimensional grid, each
program occupying one site. The size of the grid is finite, and
chosen in these experiments to be small enough that avalanches are
generally over before a new one starts. As is well-known, this is the
condition {\em sine qua non} for the observation of SOC behavior, a
separation of time scales which implies that the system is driven at
infinitesimal rates. 

Let $\tau$ denote the average duration of an
avalanche. Then, a separation of time scales occurs if the average
time between the production of new seeds of avalanches is much larger
than $\tau$. New seeds, in turn, are produced with a
frequency $\langle\epsilon\rangle P$, where $\langle\epsilon\rangle$
is again the
average replication rate, and $P$ is the mutation probability (per
replication period) for an
average sequence of length $\ell$,
\be
P=1-(1-R)^\ell\;.
\ee
For small enough $R$ and not too large $\ell$ (so that the product
$R\ell$ is smaller than unity) we can approximate
$P\approx R\ell$, and infinitesimal driving occurs in the limit 
\be
\langle \epsilon\rangle R\ell \ll\frac1\tau\;.\label{cond}
\ee
Furthermore
\be
\tau\sim\frac{L}v
\ee
with $L$ the diameter of the system and $v$ a typical Fisher velocity.
The fastest waves are those for which the latent heat is of the order
of the new fitness, i.e., $\Delta\epsilon\sim\epsilon$, in which case
$v\approx \epsilon$ (because $D\sim\epsilon$ in Eq.~(\ref{eq4}), see
Chu and Adami 1995) and 
a separation of time scales is assured whenever
\be
\frac1{R\ell}\gg {L}\;,
\ee
that is, in the limit of vanishing mutation rate or small population
sizes. For the $L=60$ system used here, this condition is obeyed (for
the fastest waves) only for the smallest mutation rate tested and
sequence lengths of the order of the ancestor.

\begin{figure}[t]
\caption{Fitness of the dominant
  genotype in the population, $\epsilon_{\rm best}$ 
  as a function of time (in updates).}
\vskip 0.25cm
\centerline{\psfig{figure=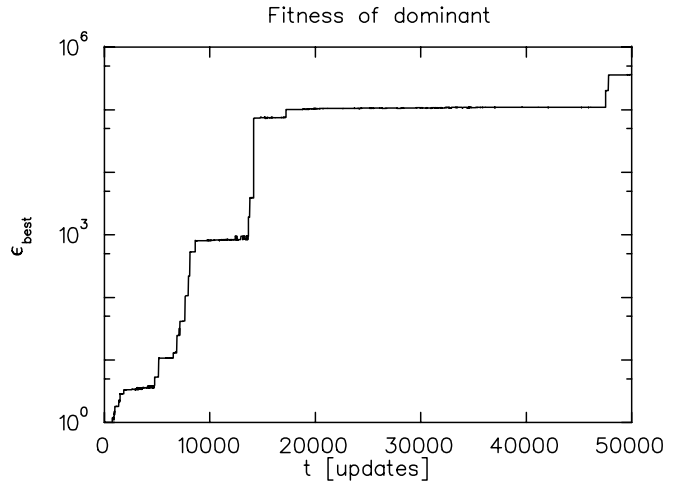,width=3.4in,angle=90}}
\label{fig2}
\end{figure}
In the following, we keep the population size constant (a $60\times60$
grid) and vary the mutation rate. From the previous arguments, we
expect true scale-free dynamics only to appear in
the limit of small mutation rates. 
As in this limit avalanches
occur less and less frequently, this is also the limit where data are
increasingly difficult to obtain, and other finite size effects can
come into play.  We shall try to isolate the scale-free regime by fitting
the distribution to a power law
\be
P(s)\sim s^{-D(R)}\label{power}
\ee
and monitor the behavior of $D$ from low to high mutation rates. 

In Fig.~\ref{fig2}, we display a typical history of $\epsilon_{\rm
  best}$, i.e., the fitness of the dominant genotype\footnote{As the
  replication rate $\epsilon$ is exponential in the bonus obtained for
  a successful computation, $\epsilon_{\rm best}$ increases
  exponentially with time.}. Note the ``staircase'' structure of the
curve reflecting the ``punctuated'' dynamics, where each step reflects
a new avalanche and concurrently an extinction event. Staircases very
much like these are also observed in adapting populations of {\it E.
  Coli} (Lenski and Travisano 1994).

As touched upon earlier, the Avida world represents an environment
replete with information, which we encode by providing bonuses for
performing logical computations on externally provided (random)
numbers. The computations rewarded usually involve two inputs $A$ and
$B$, are finite in number and listed in Table~1. At the end of a
typical run (such as Fig.~\ref{fig2}) the population of programs is
usually proficient in almost all tasks for which bonuses are given
out, and the genome length has grown to several multiples of the
initial size to accommodate the acquired information.
\begin{table}[h]
\caption{Logical calculations on random inputs $A$ and $B$ rewarded, 
bonuses, and difficulty (in minimum number of {\tt nand} instructions 
required). Bonuses $b_i$ increase the speed of a CPU by a factor 
$\nu_i=1+2^{b_i-3}$.} 
\vskip 0.25cm
\center{
\begin{tabular}{|c|c|c|c|}\hline
Name & Result & Bonus $b_i$ & Difficulty\\ \hline\hline
Echo & I/O   & 1 & --\\
Not  & $\neg A$ & 2 & 1 \\
Nand & $\neg(A\wedge B)$ & 2 & 1 \\
Not Or & $\neg A \vee B$ & 3 & 2 \\
And  &  $ A \wedge B $   & 3 & 2 \\
Or   &  $ A \vee B $     & 4 & 3 \\
And Not & $A\wedge\neg B$& 4 & 3 \\
Nor  & $\neg(A\vee B)$   & 5 & 4 \\
Xor  & $ A\ {\rm xor}\ B$ &   6 & 4 \\
Equals &$\neg(A\ {\rm xor}\ B)$&6& 4 \\ \hline
\end{tabular}  
}
\end{table}

\begin{figure}[h]
\caption{Distribution of genotypes sizes $P(s)$ fitted to a power law
  (solid line) at mutation rate $R=0.004$.} 
\vskip 0.25cm
\centerline{\psfig{figure=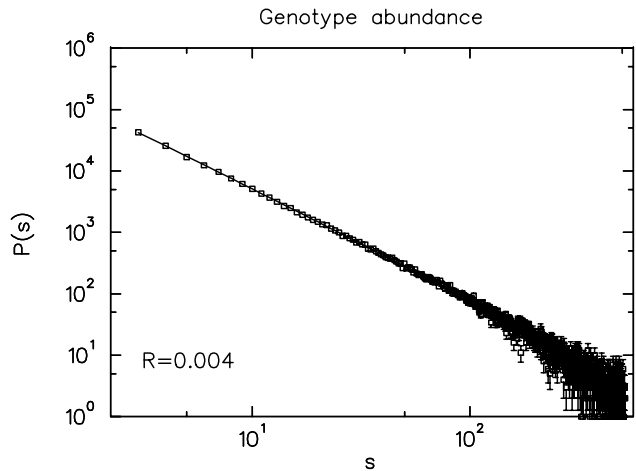,width=3.25in,angle=90}}
\label{fig3}
\end{figure}

\begin{figure}[!]
\caption{Distribution of genotypes sizes $P(s)$ for a landscape devoid
  of the bonuses listed in Tab.~1, at mutation rate $R=0.003$.} 
\vskip 0.25cm
\centerline{\psfig{figure=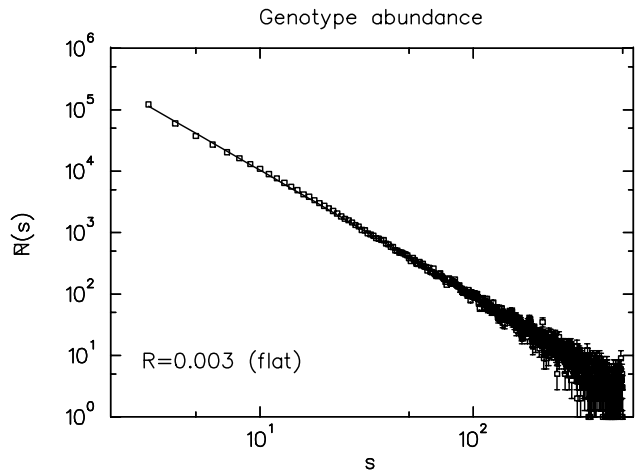,width=3.25in,angle=90}}
\label{fig4}
\end{figure}

Because the amount of information stored in the landscape is finite, 
adaptation, and the associated avalanches,
must stop when the population has exhausted the landscape. 
However, we shall see that even a `flat' landscape (on which evolution
is essentially neutral after the sequence has optimized its
replicative strategy ) gives rise to a power law of genotype sizes, as long as
the programs do not harbor an excessive amount of ``junk''
instructions\footnote{
``Junk'' instructions do not code for any
  information, and do not affect the fitness of their
  bearer. Consequently, programs with excessive amounts of junk code
  will give rise to many ``degenerate'' genotypes with no competitive
  advantage. In this regime, the genotype abundance distribution is
  exponential rather than of the power-law type, due to a violation of
  condition (\ref{cond}).}.
A typical abundance distribution (for the run 
depicted in Fig.~\ref{fig2}) is shown in Fig.~\ref{fig3}. 
As mentioned earlier, we can also
turn {\em off} all bonuses listed in Tab.~1, in which case fitness is
related to replicative abilities only. Still, avalanches occur
(within the first 50,000 updates monitored) due to minute improvements
in fitness, but the length of the genomes typically stays in the range
of the ancestor, a program of length 31 instructions. We expect a
change of dynamics once the ``true'' maximum of the local fitness
landscape is reached, however, we did not reach this regime in the
experiments presented here. The distribution of genotype sizes for the
flat landscape is depicted in Fig.~\ref{fig4}.
Clearly then, even such landscapes (flat with respect to all other
activities except replication) are not neutral. Indeed, it is known
that neutral evolution, where the chance for a genotype to increase or
decrease in number is even, leads to a power law in the abundance
distribution with exponent $D=1.5$ (Adami, Brown, and Haggerty, 1995).   

\begin{figure}[t]
\caption{Fitted exponent of power law for 34 runs at mutation rates
  between $R=0.0005$ and $R=0.01$ copy errors per instruction
  copied. The error bars reflect the standard deviation across the
  sample of runs taken at each mutation rate. The solid line is to
  guide the eye only. 
}
\vskip 0.25cm
\centerline{\psfig{figure=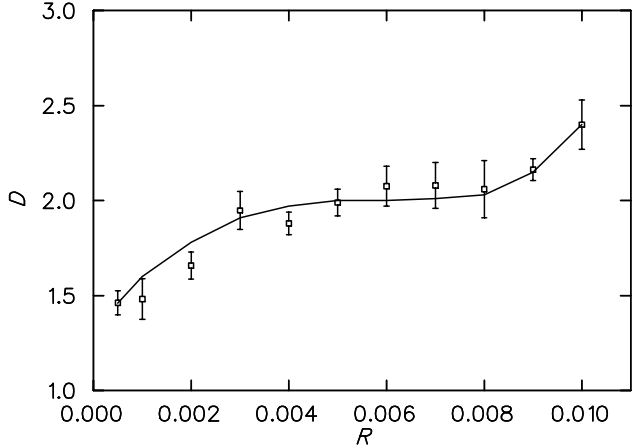,width=3.25in,angle=90}}
\label{fig5}
\end{figure}

In order to test the dependence of the fitted exponent $D(R)$
[Eq.~(\ref{power})] on 
the mutation rate, we conduct a set of experiments at varying
copy-mutation rates from $0.5\times10^{-3}$ to $10\times10^{-3}$ and 
take data for 50,000 updates. Again, a ``best'' genotype is not
reached after this time, and we must assume that avalanches were still
occurring at the end of these runs. Furthermore, in some runs we find
that a genotype comes to dominate the population (usually after most
`genes' have been discovered) which carries an unusual amount of junk
instructions. As mentioned earlier, such species produce a
distribution that is exponentially suppressed at large genotype sizes
(data not shown). To avoid contamination from such species, we stop
recording genotypes after a plateau of fitness was reached, i.e., if
the population had discovered most of the bonuses. Furthermore, in
order to minimize finite
size effects on the determination of the critical exponent, we
excluded from this fit all genotype abundances larger than 15, i.e.,
we only fitted the smallest abundances. Indeed, at larger mutation
rates the higher abundances are contaminated by a pile-up effect due to
the toroidal geometry, while at lower mutation rates a scale appears
to enter which prevents scale-free behavior. We have not, as yet,
been able to determine the origin of this scale.

In the results
reported here, we show the dependence of the fitted exponent $D$ as a
function of the mutation rate $R$ used in the run, which, however, is a good
measure of the mutation probability $P$ only at small $R$ and if the
sequence length is not excessive. As a consequence, data points at
large $R$, as well as runs where an excessive sequence length
developed, carry a systematic error. 

For the 34 runs that we obtained, the power $D$ was measured for each
run (for the low abundances), and an average was calculated for all
the runs at a particular mutation rate. This data is plotted in
Fig.~\ref{fig5} and shows a plateau in the fitted exponent only
at intermediate mutation rates, with $D=2.0\pm 0.05$. A fit of the
middle abundances (10-100) produces a critical coefficient more or
less independent of mutation rate, around $D=2.0$, but with less
accuracy (data not shown). At high $R$, we witness a deviation from
scale-free behavior (reflected in the rising $D$ for small abundances)
which is most likely due to pile-up, i.e., a finite toroidal lattice.
This effect may be avoided by using absorbing rather than periodic
boundary conditions. We also see a violation of scaling at small $R$,
which is due to the emergence of some other scale. While it is most
likely a finite-size effect, the exact origin of this scale is as yet
unclear.  We comment on the significance of these results in Section V.

Still, more control over the
spread in exponents for fixed mutation rate would be desirable. This
can obviously be achieved by plotting $D$ versus $P$, rather than $R$,
for example, and by
better keeping track of the coding percentage within a genotype, a
variable that we know significantly affects the shape of the
distribution. Such experiments are planned for the near future.     
\section{Distribution of Species Sizes}
In Avida, it is possible to monitor groups of programs that display
the same ``phenotype'', while differing in genotype. Even though
programs in this world are haploid (single-stranded) and do not
reproduce sexually, it is convenient to label such groups
taxonomically, i.e., we refer to them as ``species''. Strictly
speaking, a species consists out of  all those genotypes that, when
executed, give rise to the same ``chemistry'', i.e., such programs
differ only in instructions that are either unexecuted, or else are
{\em neutral}. Algorithmically, the determination whether two
genotypes belong to the same species is complicated by the fact that
sequence length is {\em not} constant in these experiments. Thus, we
need to be able to compare strings with differing lengths,
which is achieved by lining them up in such a manner that they are
identical in the maximum number of corresponding sites. Subsequently,
a {\em cross-over} point is chosen randomly and the genomes above and below
this point are {\em swapped}. In other words, we construct a {\em
  hybrid} program from the two candidates and test it for
functionality, but without introducing it in the population
(see Adami 1998.) In the experiments reported here, we actually test
{\em two} cross-over points in order to rule out accidental
matches. In retrospect, we find that almost all those strings
classified as belonging to the same species by this method differ only
in ``silent'', or at least inconsequential,  instructions. 

The abundance distribution of genotypes within species more closely
corresponds to the kind of geometric distributions investigated by 
Willis (1922) as well as Burlando (1990, 1993). Indeed, Burlando
found, in an analysis of distributions of subtaxa within taxa obtained
from the fossil record as well as recorded flora and fauna, that
these distributions appear to be scale-free across taxonomic hierarchies,
with critical coefficients between $2.0 < D < 2.5$. This distribution
can also be viewed as a distribution of avalanches sizes, if
avalanches are redefined as events that spawn different genotypes of
the same species. Indeed, in this manner it is possible to investigate
hierarchies of avalanches, each higher level presumably sporting a
higher critical coefficient.

In the
experiments reported here, we found species coefficients closer to
$D\approx 2.5$, but we also found violations of power-law behavior
which are most likely due to the contribution of species of different
lengths to the abundance distribution. Indeed, the amount of ``junk''
instructions in a species most likely governs the steepness of the
distribution, and several different such species may give rise to a
{\em multifractal} distribution rather than a pure power law. In the
future, we expect to disentangle such distributions by appealing to a an even
higher level in taxonomy, reuniting all species of the same sequence
length within a {\em genus}. The latter taxonomic level could, for
example, be entirely phenotypic, by keeping track of which tasks a
genus executes (irrespective of its genotype).

Still, even though changing sequence lengths affect the distribution
of genotypes within species, those experiments in which the sequence
length does {\em not} change significantly can give rise to power laws with
single exponents, as shown below in Fig.~\ref{fig6}. The data for this
experiment were obtained from the same run as gave rise to
Figs.~\ref{fig2} and \ref{fig3}. 
\begin{figure}[t]
\caption{Distribution of genotypes within species at $R=0.004$, fitted
  to a power law with $D=2.44\pm0.05$.}
\vskip 0.25cm
\centerline{\psfig{figure=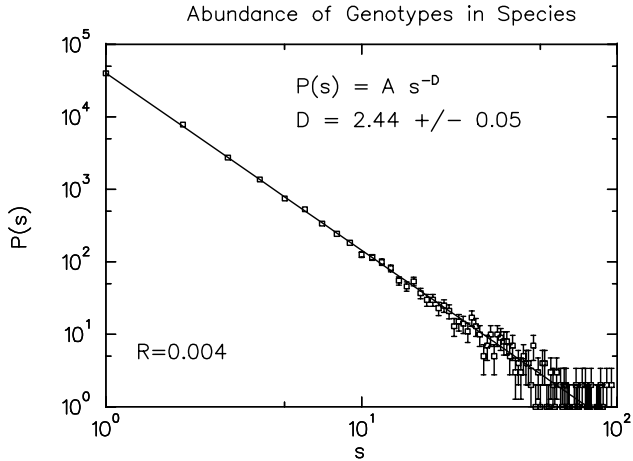,width=3.25in,angle=90}}
\label{fig6}
\end{figure}
\section{Conclusions}
The distribution of avalanche sizes in evolving systems, which is
quite clearly related to 
the distribution of extinction events, can reveal a fair amount of
information about the dynamics of the adapting agents. For example,
purely random systems in which there are no fitness advantages, and where 
selection does not occur, can still show power law behavior, as
extinction events are governed by the return-to-zero probability of
random walks\nocite{ABH} (Adami, Brown and Haggerty 1995). 
In Avida, we observe a scaling exponent $D=2.0$ in an intermediate
regime of mutation rates. While it is still unclear whether the {\em
  mixing of scales} that we have observed at small and large mutation
rates is due to the finite size of the lattice or the emergence of 
another scale, we can conclude with confidence that scale-free dynamics
does occur. Scaling violations should be investigated by a thorough
finite lattice-size analysis, and this is planned for the future along
with more refined methods for dealing with explicit neutrality (i.e.,
``junk'' code.) 

An interesting hint at what the distribution might be like in Nature
comes from Raup's analysis of a data set prepared by
Sepkoski\nocite{RAU} (Raup 1991): genera of marine invertebrates from
the fossil record. Raup's ``kill-curve'' can be transformed into a
distribution of sizes of extinction evens (as shown by
Newman\nocite{NEW1} 1996) governed by a critical exponent close to
$D=2.0$. This is tantalizingly close to the coefficient we found in
our genotype abundance distribution, but we must be careful in
comparing these distributions.

The avalanche-size distribution of genotypes gives us a good
indication of the strength of an evolutionary shock, but also about
the length of time the particular species dominates the dynamics, and
therefore, of the time {\em between} evolutionary transitions. Also, each
evolutionary transition brings with it a wave of extinction, as all
previously extant genotypes and species of lower fitness must
disappear on the heels of the new ``discovery''. The size of
extinction events proper, however, is not measured by the
``epoch-length'' distribution reflected in the avalanche sizes, but
rather by the abundance of genotypes within species (or any higher
taxonomic abundance distribution) because each species appearing in
this distribution must eventually go extinct, and thus this
distribution must equal the distribution of extinction sizes. The
latter distribution (measured in Section IV), appears to have a
critical exponent around $D\approx 2.5$, higher than the corresponding
one from the fossil record. Furthermore, we must keep in mind the
simplicity of the model treated here when comparing to actual fossil
data. As mentioned in the introduction, coevolution does not play a
role in the dynamics controlling the size of avalanches in this model,
while we must assume that extinctions in Earth history have some
co-evolutionary component. On the other hand, the abundance
distribution of genotypes within species is consistent with those
obtained by Burlando (1990, 1993), who argued that they represented
evidence for a ``fractal geometry of Nature''.
 
From the present analysis, it is clear that there is as yet no reason
to jump to conclusions from the evidence extracted either from the
fossil record, theoretical models of extinctions\nocite{NEW2} (Newman
1997), or else direct implementation of the dynamics of adaptive
avalanches as we have done here. We do, however, see clear evidence
that avalanches not reigned in by any scale can and do develop in
evolving and adapting systems {\em without} co-evolutionary pressures,
via first-order transitions in populations occupying single ecological
niches. Not only do we find scale-free dynamics for the time between
transitions (as evidenced by the genotype abundance distribution) but
also for the {\em strength} of these transitions, measured by the
distribution of species-sizes.  It is left for future experiments to
determine how such dynamics, taking place in {\em interacting}
ecological niches, gives rise to power laws for co-evolutionary
systems, and how the description in terms of first-order transitions
is {\it ipso facto} transmutated into a second-order scenario.

This work was supported by NSF grant No.\ PHY-9723972.


\begin{thebibliography}{99}
\bibitem{CA2}Adami, C. 1995. Self-organized criticality in living
  systems. {\em Phys. Lett.} A 203:~23.
\bibitem{IAL}Adami, C. 1998. {\it Introduction to Artificial
  Life}. Santa Clara: TELOS Springer-Verlag.
\bibitem{AB1} Adami, C. and C. T. Brown. 1994. Evolutionary learning in
  the 2D Artificial Life system `Avida'. In {\em Artificial Life
  IV}, edited by R.A. Brooks and P. Maes. Cambridge, MA: MIT Press, p. 377.
\bibitem{ABH}Adami, C., C. T. Brown and M. R. Haggerty. 1995. Abundance
  distributions in Artificial Life and stochastic models: `Age and
  Area' revisited.{\em  Lect. Notes in Artif. Intell.} 929:~503. 
\bibitem{BP96}Bak, P. and M. Paczuski. 1996. In {\it Physics of Biological
    Systems}. Heidelberg: Springer-Verlag. 
\bibitem{BS}Bak, P. and K. Sneppen. 1993. Punctuated equilibrium and
  criticality in a simple model of evolution. {\em Phys. Rev. Lett.} 71:~4083.
\bibitem[Bak, Tang, and Wiesenfeld, 1987]{BTW87} 
Bak, B., C. Tang, and K. Wiesenfeld. 1987. Self-organized criticality:
  An explanation of 1/$f$ noise. {\em Phys. Rev. Lett.} 59:~381.
\bibitem[Bak, Tang, and Wiesenfeld, 1988]{BTW88}
Bak, B., C. Tang, and K. Wiesenfeld. 1988. Self-organized criticality.
{\em Phys. Rev.} A 38:~364.
\bibitem{BUR90} Burlando, B. 1990. The fractal dimension of taxonomic
  systems. {\em J. Theor. Biol.} 146: 99.
\bibitem{BUR93} Burlando, B. 1993. The fractal geometry of evolution.
{\em J. Theor. Biol.} 163:~161.
\bibitem{CHU}Chu, J. and C. Adami. 1997. Propagation of information in
  populations of self-replicating code. In Proc. of {\it Artificial Life V},
edited by C. G. Langton and K. Shimohara. Cambridge, MA: MIT Press, p. 462. 
\bibitem{ECL96}Elena, S. F., V. S. Cooper and
  R.E. Lenski. 1996. Punctuated evolution caused by selection of rare
  beneficial mutations. {\em Science} 272:~1802. 
\bibitem{GS}Gil, L. and D. Sornette. 1996. Landau-Ginzburg theory of
  self-organized criticality. {\em Phys. Rev. Lett.} 76:~3991.
\bibitem{GE77}Gould, S. J. and N. Eldredge. 1977. Punctuated
  equilibria: The tempo and mode of evolution reconsidered. {\em
  Paleobiology} 3:~115.
\bibitem{GE93}Gould, S. J. and N. Eldredge. 1993. Punctuated
  equilibrium comes of age. {\em Nature} 366:~223.
\bibitem{LT94}Lenski, R. and M. Travisano. 1994. Dynamics of
  adaptation and diversification--A 10,000 generation experiment with
  bacterial populations. {\em Proc. Nat. Acad. Sci.} 91:~6808-6814.
\bibitem{NEW1}Newman, M. E. J. 1996. Self-organized criticality,
  evolution, and the fossil extinction record. {\em Proc. Roy. Soc.} 
B 263:~1605--1610.
\bibitem{NEW2}Newman, M. E. J. 1997. A model of mass extinction. Eprint
  adap-org/9702003. 
\bibitem{NFST}Newman, M. E. J., S. M. Fraser, K. Sneppen, and
  W.A. Tozier. 1997. Self-organized criticality in living
  systems---Comment. {\em Phys. Lett.} A 228:~202. 
\bibitem{OBA}Ofria, C., C. T. Brown and C. Adami. 1998. {\it The Avida User's
    Manual}. In Adami (1998). The Avida software is publicly available
  at ftp.krl.caltech.edu/pub/avida.
\bibitem{RAU}Raup, D. M. 1991. A kill curve for phanerozoic marine species.
{\em Paleobiology} 17:~37--48..
\bibitem{SJD}Sornette, D., A. Johansen, and I. Dornic. 1995. Mapping
  self-organized criticality to criticality. {\em J. de Phys.}
  I 5:~325. 
\bibitem{SB96}Sol\'e, R. V. and J. Bascompte. 1996. Are critical
  phenomena relevant to large-scale evolution? {\em Proc. Roy. Soc.} B
  263:~161--168. 
\bibitem{WIL}Willis, J. C. 1922. {\em Age and Area}. Cambridge:
  Cambridge University Press.
%
\end{thebibliography}
\end{document}